\documentclass[10pt,aps,prb,twocolumn,amssymb,footinbib,letterpaper,superscriptaddress]{revtex4-1}
\usepackage{graphicx}
\usepackage{color}
\usepackage{dcolumn} % Align table columns on decimal point
\usepackage{bm}      % bold math
\usepackage{array}   % table making
\usepackage{amssymb}
\usepackage{amsfonts}
\usepackage{amsmath}
\usepackage[colorlinks,citecolor=blue]{hyperref}
\newcommand{\EvRes}{ E_{var}^{Res}}
\newcommand{\EvFED}{ E_{var}^{Fed}}
\newcommand{\EcRes}{ E_{qmc}^{Res}}
\newcommand{\EcFED}{ E_{qmc}^{Fed}}
\newcommand{\dqRes}{\Delta_{qmc}^{Res} }
\newcommand{\dqFED}{\Delta_{qmc}^{Fed} }
\newcommand{\dvRes}{\Delta_{var}^{Res} }
\newcommand{\dvFED}{\Delta_{var}^{Fed} }
\newcommand{\conf}{(N_\upa, N_\dna)}
\newcommand{\sysa}{{\mbox{sqr}\,\,4\times 4}}
\newcommand{\sysb}{{\mbox{hc}\,\,2\times 3}}
\newcommand{\sysc}{{\mbox{tri}\,\,4\times 3}}
\newcommand{\sysd}{{\mbox{tri}\,\,4\times 4}}
\newcommand{\syse}{{\mbox{kag}\,\,2\times 2}}

\newcommand{\dtau}{\Delta\tau}

\newcolumntype{.}[1]{D{.}{.}{#1}}

\newcommand{\be}{\begin{equation}}
\newcommand{\ee}{\end{equation}}
\newcommand{\ba}{\begin{eqnarray}}
\newcommand{\ea}{\end{eqnarray}}

\newcommand{\upa}{\uparrow}
\newcommand{\dna}{\downarrow}

\newcommand{\COMMENTED}[1]{}

\newcommand{\ket}[1]{|#1\rangle}

\newcommand{\ca}[2]{{\hat c}_{#1 #2}^\dagger}
\newcommand{\de}[2]{\hat{c}_{#1 #2}^{\phantom{\dagger}}}

\newcommand{\ob}[1]{{\langle #1\rangle}}

\newcommand{\bfa}{\mathbf{a}}

\newcommand{\bfi}{\mathbf{i}}
\newcommand{\bfj}{\mathbf{j}}

\newcommand{\bfs}{\mathbf{s}}

\newcommand{\hB}{{\hat{B}}}

\newcommand{\hH}{{\hat{H}}}

\newcommand{\hK}{{\hat{K}}}

\newcommand{\hV}{{\hat{V}}}

\newcommand{\hc}{{\hat{c}}}

\newcommand{\hn}{{\hat{n}}}

\newcommand{\hv}{{\hat{v}}}

\begin{document}

\title{Multi-determinant generalized Hartree-Fock wave functions in Monte Carlo calculations}

\author{Chia-Chen Chang}
\affiliation{Department of Physics, University of California Davis, CA 95616, USA}
\author{Miguel A. Morales}  
\affiliation{Physics Division, Lawrence Livermore National Laboratory, Livermore, CA 94550, USA}

\begin{abstract}

The quantum Monte Carlo algorithm is arguably one of the most powerful computational 
many-body methods, enabling accurate calculation of many properties in interacting 
quantum systems. In the presence of the so-called sign problem, the algorithm typically 
relies on trial wave functions to eliminate the exponential decay of signal-to-noise ratio, 
usually at the expense of a bias in the result. The quality of the trial state therefore 
is critical for accurate simulations. In this work, benchmark results of the ground state 
auxiliary-field quantum Monte Carlo method are reported for the Hubbard model on several 
geometries. 
We demonstrate that when multi-determinant generalized Hartree-Fock states are used as 
trial wave functions, the systematic errors can be systematically reduced to a low level 
and the results compare favorably with the exact diagonalization data. 

\end{abstract}

\maketitle

% --------------------------------------------------------------------------------------
\section{Introduction}

Strongly correlated materials, for example cuprate superconductors\cite{Bednorz:1986} and heavy 
fermions,\cite{Coleman:2007,Si:2010} host a variety of remarkable phenomena such as high-$T_c$ 
superconductivity,\cite{Lee:2006,Keimer:2015} metal-to-insulator transition,\cite{Imada:1998} 
magnetism,\cite{Sachdev:2008} quantum criticality,\cite{Gegenwart:2008} to name a few. 
The rich phases of these correlated materials offer great potential for energy and technology 
applications. Understanding and predicting their electronic and structural properties is one of 
the central missions of modern condensed matter physics and quantum chemistry. Nonetheless, the 
strongly correlated nature and complexity of the materials often pose great challenges to traditional 
theoretical methods. With the advances in hardware and algorithms, computational many-body techniques 
have become increasingly important in unlocking the underlying mechanisms of the physics emerging 
from strong correlations.

Over the past few decades, various numerical tools have been devised to solve the many-body
Schr\"odinger equation for models as well as materials. Exact methods such as direct diagonalization 
and configuration interaction\cite{Sherrill:1999} provide the most unbiased standard. However, 
the computational cost scales exponentially with system size since the methods deal directly 
with the Hilbert space of the Hamiltonian.
Density functional theory (DFT),\cite{Jones:2015} on the other hand, drastically reduces the 
computational cost by mapping the many-body problem onto an effective single-electron one 
through the use functionals that depend on the electron density. For its ability to handle 
large complex structures at a rather low computational cost, DFT has been widely used in the 
field of condensed matter physics, quantum chemistry, and material sciences. In spite of its 
success, limitations of DFT exist and treating strongly correlated systems with DFT remains 
challenging.\cite{Cohen:2012} It should be pointed out that progress has been made by combining 
DFT with other many-body techniques such as density-matrix renormalization 
group\cite{Stoudenmire:2012} or dynamical mean-field theory.\cite{Anisimov:1997,Lichtenstein:1998}

In this regard, quantum Monte Carlo (QMC) methods offer an alternative route for tackling the 
problem of strong correlations. On the one hand, QMC methods treat electronic correlations 
directly without the need for empirical or approximate interactions. On the other hand, it 
offers a favorable algebraic scaling (${\cal O}(N^3) \sim {\cal O}(N^4)$ where $N$ is the 
system size) compared to exact methods that typically scale exponentially with system size. 
For this reason, QMC methods can handle fairly large system sizes and provide near exact 
answers with controllable statistical errors. For example, in the study of helium-4 atoms 
under extreme pressure or temperature conditions, path integral Monte Carlo has been the 
choice of method.\cite{Ceperley:1995} 

For fermionic systems, however, the probability distribution used to sample the Hilbert space 
is no longer positive-definite due to the Pauli exclusion principle. Under this circumstance, 
statistical error grows exponentially and QMC loses its algebraic scaling, leading to the so-called 
sign problem.\cite{Troyer:2005,Iazzi:2016} In most cases, there are approximate solutions to 
the sign problem which control the exponential decay of the signal-to-noise ratio. In the well-known 
fixed-node diffusion Monte Carlo (FNDMC) method,\cite{Foulkes:2001} for example, a trial wave 
function is used to eliminate any moves that try to cross the nodal hyper-surface of the reference. 
In the same spirit, the ground state auxiliary-field quantum Monte Carlo (AFQMC) method implements 
the constrained-path approximation\cite{Zhang1995,Zhang1997} to retain positive random walkers. 
Under these approximations both QMC techniques regain the algebraic scaling, at the cost of a bias
in the simulation results that depend sensitively on the quality of the trial wave function used 
to implement the constraint. Consequently, accurate trial wave functions are vital for the efficiency 
and correctness of simulations. In this work, we propose a class of multi-determinant mean-field 
trial wave functions for the AFQMC method. By comparing simulation results with exact diagonalization
data, it is shown that the variational freedom of the proposed multi-determinant trial wave functions 
allow one to deliver accurate simulations. 

The rest of the paper is organized as follows. Section \ref{sec:Method} briefly summarizes the 
algorithms adopted to prepare the trial wave functions and the constrained-path AFQMC method. 
Benchmark results and discussions are presented in Section \ref{sec:RD}. The paper will be closed 
with a short summary.

% --------------------------------------------------------------------------------------
\section{Method}
\label{sec:Method}
% --------------------------------------------------------------------------------------

% --------------------------------------------------------------------------------------
\subsection{Multi-determinant generalized Hartree-Fock wave function}

In this work we employ trial wave-functions consisting of non-orthogonal Slater determinant 
expansions of the form:
\be
  \ket{\Phi} = \sum_{i=1}^{n_d}\,c_i\,\ket{\varphi_i},
  \label{eq:FEDwf}
\ee
where $\ket{\varphi_i}$ are Slater determinants and $c_i$ are linear variational parameters. 
As opposed to Slater determinant expansions built from particle-hole excitations from a given 
reference determinant, typically employed in quantum Monte Carlo calculations, we do not enforce 
orthogonalization between different Slater determinants in the expansion, hence 
$\ob{\varphi_i|\varphi_j}\neq0$. In fact, each Slater determinant is represented
as an orbital rotation from a given reference
\be
\ket{\varphi_i} = e^{\sum_{pq} Z^i_{pq} \ca{p}{}\de{q}{}} \ket{\varphi_{\mbox{\small ref}}}, 
\ee
where $Z$ is an unitary matrix. 

The trial wave function is generated using a version of the projected Hartree-Fock (PHF) algorithm
developed in the Scuseria group at Rice University.\cite{Jimenez:2012a,*JimenezHoyos:2013,*Schutski:2014}
The algorithm directly minimizes the energy $E=\ob{\Phi|\hH|\Phi}/\ob{\Phi|\Phi}$ using a BFGS-like 
technique and analytical energy gradients. We refer the readers to Ref.~\onlinecite{Jimenez:2012a,
*JimenezHoyos:2013,*Schutski:2014} 
for the relevant equations. There are two different approaches within the algorithm, 
the few-determinant (FED) algorithm\cite{RodriguezGuzman:2013,RodriguezGuzman:2014} and 
the resonating Hartree-Fock (ResHF) approach.\cite{Fukutome:1988,Yamamoto:1991,Tomita:2004}
In the FED algorithm, the Slater determinant expansion is generated iteratively, adding and optimizing 
one determinant in each iteration to an already existing expansion. During each iteration, determinants 
$\ket{\varphi_i}$  ($i=1,2,\ldots, n_d-1$) obtained from previous iterations are kept 
fixed\cite{RodriguezGuzman:2013,RodriguezGuzman:2014} and the energy is minimized with respect to 
the orbital rotation matrix of the new determinant and all linear coefficients. This process continues 
until a given number of determinants is generated. At this point, the linear coefficients are 
re-optimized by solving the associated eigenvalue problem. In the FED theory, symmetry projectors 
can be incorporated straightforwardly. The resulting single- or multi-reference symmetry-projected 
FED wave functions have been shown to be quite accurate.\cite{Shi:2013jj,Shi:2014bw} 
However, we will not focus on symmetry restoration in our calculations.

In the ResHF approach the energy, $E$, is minimized with respect to all variational parameters in 
the trial wave-function, including the rotation matrices of all determinants and all linear coefficients. 
In this work, ResHF trial wave-functions are produced using FED generated trial wave-functions as input. 
We use the same BFGS-like direct optimization algorithm used for FED, but include all parameters
simultaneously in the optimization. 

The mean-field orbital used in the FED and ResHF theories could be restricted-HF (RHF), unrestricted-HF (UHF),
or generalized HF (GHF) wave functions. The three types of state represent different levels of symmetry-breaking. 
In the present study, we will mainly focus on using GHF determinants in which all symmetries of the Hamiltonian are
broken except for the total particle number. However, it is argued that the GHF wave function contains the most 
variational freedom (at the mean-field level)\cite{Hammes-Schiffer:1993} comparing to the RHF or UHF states. 
In the following, we will demonstrate the flexibility of the symmetry-broken multi-determinant GHF wave function 
in QMC calculations through the combined strength of the FED and ResHF theories.

\subsection{AFQMC and the Constrained-Path approximation}

The AFQMC technique implemented in this study works in the second-quantized framework. It is based on the 
projection equation
\be
  \ket{\Psi_0} = \lim_{\tau\rightarrow\infty} e^{-\tau\hH}\ket{\Psi_T},
\ee
where $\tau$ is the imaginary time and $\ket{\Psi_T}$ is a trial wave function that has a finite
overlap with the many-body ground state $\ket{\Psi_0}$ of the Hamiltonian $\hH$. In actual numerical
calculations, the projection is realized iteratively
\be
  \ket{\Psi^{(\ell+1)}} = e^{-\dtau\hH}\ket{\Psi^{(\ell)}},
\ee
where the discrete imaginary time step $\dtau$ is chosen to be a small positive number.

In a numerical calculation, the small-time projector is first approximated by 
implementing the second-order Trotter-Suzuki break-up\cite{Trotter:1959,Suzuki:1976}
\be
  e^{-\dtau\hH} \approx e^{-\dtau\hK/2}e^{-\dtau\hV}e^{-\dtau\hK/2},
  \label{eq:Trotter}
\ee
where $\hK$ and $\hV$ denote the part of $\hH$ that is quadratic and quartic in fermion operators
respectively. The error of the approximation is of the order ${\cal O}(\dtau^2)$ and can be reduced 
systematically using higher order break-up formulas. 
The quartic term is further decomposed using the Hubbard-Stratonovich (HS) transformation\cite{Hubbard:1959}
\be
  e^{-\dtau\hV} = \int d\bfs\,P(\bfs)\,e^{\hv(\bfs)}.
  \label{eq:HStransformation}
\ee
In this identity, $P(\bfs)$ is a probability distribution function of the HS fields $\bfs$ and its 
functional form is determined by the quartic term $\hV$ of the Hamiltonian. $\hv(\bfs)$ is an one-body 
operator that depends on $\bfs$, $\dtau$, and the matrix elements of $\hV$ (only the $\bfs$-dependence 
is shown explicitly).
 
Using Eqs.~(\ref{eq:Trotter}) and (\ref{eq:HStransformation}), the iterative projection equation can 
be cast as 
\be
  \ket{\Psi^{(\ell+1)}} = \int d\bfs\,P(\bfs)\,\hB(\bfs)\,\ket{\Psi^{(\ell)}},
\ee
where we have used $\hB(\bfs)$ to denote collectively the product of one-body projectors 
$e^{-\dtau\hK/2}e^{\hv(\bfs)}e^{-\dtau\hK/2}$. The integral equation is then realized by
importance-sampled branching random walks in the space of Slater determinants.
At every step, $\hB(\bfs)$ is applied to each walker $\ket{\phi^{(\ell)}}$ in a population
by drawing a component of the HS field $\bfs$ from the distribution $P(\bfs)$. This process
generates a new collection of walkers for the next iteration. The walker population varies
along the way of projection due to fluctuating weights. Once converged, the resulting 
walkers, together with their weights, give a stochastic representation of the ground state 
wave function $\ket{\Psi_0}$ at any step. Observables such as energy and correlation 
functions can be measured periodically.

Within the AFQMC approach just described, the sign problem occurs because of the symmetry between 
the walkers with opposite overall signs $\{\ket{\phi}\}$ and $\{-\ket{\phi}\}$.\cite{Zhang1997} The two 
sets are degenerate and cannot be distinguished by the random walks. If uncontrolled along the 
projection, the Monte Carlo representation of the ground state wave function will eventually 
become an equal mixture of $+$ and $-$ walkers. The overall Monte Carlo signal therefore decays 
(exponentially fast) with imaginary time, leading to the sign problem.\footnote{In certain cases,
such as the half-filled repulsive single-band Hubbard model on bipartite lattices, there is no 
sign problem due to the particle-hole symmetry of the Hamiltonian.} The sign problem can be 
eliminated by removing walkers that have zero overlap with the ground state, $\ob{\Psi_0|\phi}=0$,
because these walkers will no longer contribute to the representation of the ground state in the 
future.\cite{Zhang1997} Since the exact ground state $\ket{\Psi_0}$ is typically unknown, the 
constrained-path approximation\cite{Zhang1995,Zhang1997} uses the trial wave function to enforce 
the constraint, $\ob{\Psi_T|\phi}=0$, at each step of the projection. As a consequence, the AFQMC
algorithm under the constrained-path approximation (the combined technique will be called CPMC
in the following discussions) is not exact and the results
will have a systematic error that depend on the quality of the trial wave function. 
Previous benchmarks for the single-band repulsive Hubbard model have shown that the constrained-path 
error is typically very small with single-determinant mean-field type trial wave 
functions.\cite{Zhang1997,Shi:2013jj,Shi:2014bw,Qin:2016,Chang:2016} While there
are proposals for more elaborate trial states which give accurate results\cite{Shi:2013jj,Shi:2014bw,Chang:2016},
we will show below that the symmetry-broken MR GHF state provides a robust
and computationally economic route for accurate CPMC calculations.

% ------------------- figure ----------------------
\begin{figure}[b]
\includegraphics[scale=0.42]{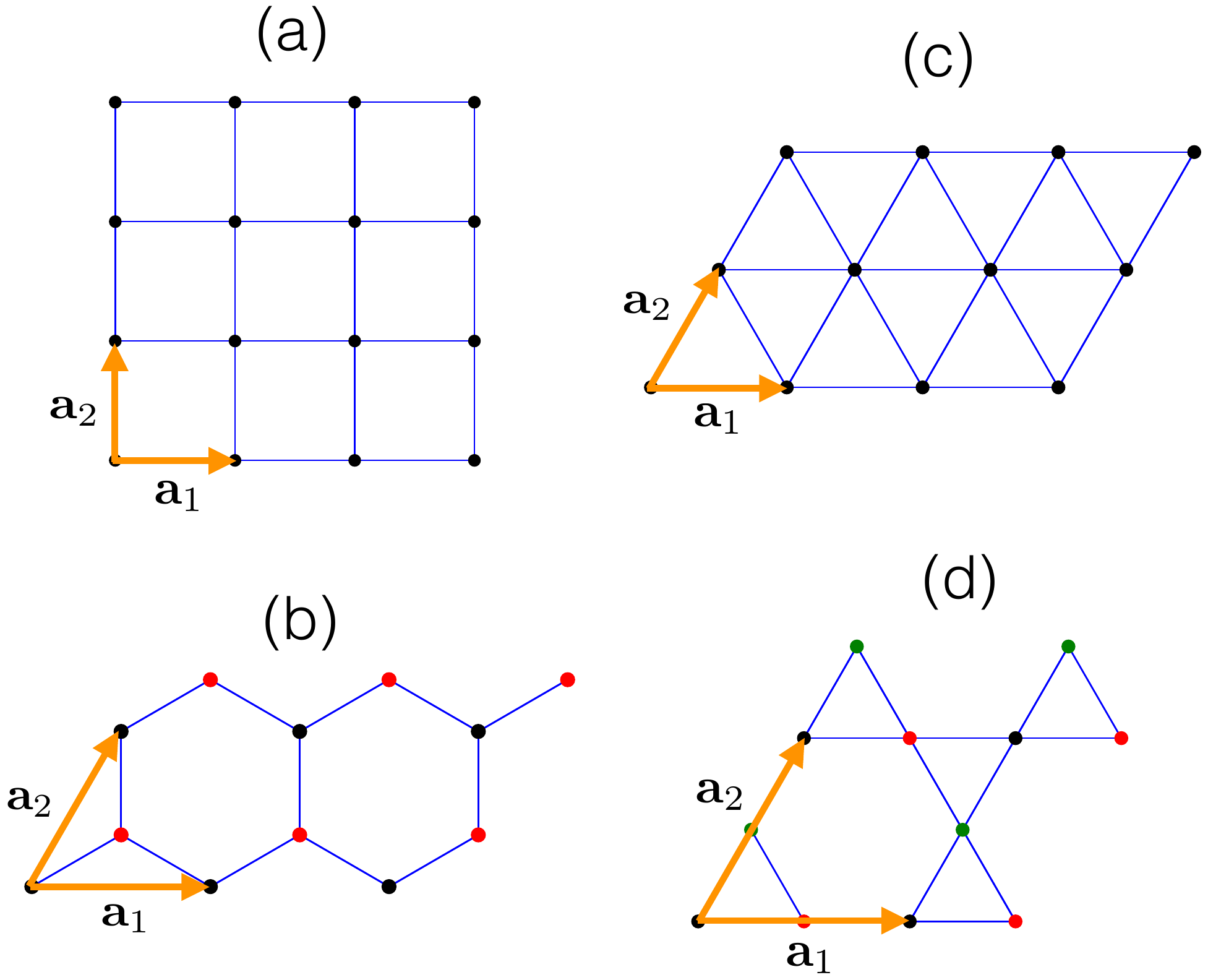}
\caption{Geometries considered in this work: (a) $4\times 4$ square lattice,
(b) $3\times 2$ honeycomb lattice, (c) $4\times 3$ triangular lattice,
and (d) $2\times 2$ kagome lattice. In each panel, the colors represent 
different basis in a unit cell. $\bfa_1$ and $\bfa_2$ are Bravais lattice vectors.
The square lattice has $\bfa_1=(a,0)$ and $\bfa_2=(0,a)$, while the lattice
vectors are $\bfa_1=(a,0)$ and $\bfa_2=a(\frac 1 2, \frac{\sqrt{3}}{2})$ for the
rest of the geometries. We set the lattice constant $a=1$.
}
\label{fig:lattices}
\end{figure}

\section{Results and Discussion}
\label{sec:RD}

\subsection{Benchmark platform}

We choose to demonstrate the symmetry-broken multi-determinant GHF state by solving the ground state of the 
repulsive Hubbard model using the CPMC algorithm. The model is defined by the Hamiltonian
\begin{align}
\hH = & -t \sum_{\left\langle \bfi\bfj\right\rangle \sigma} 
       \left( \hc^{\dagger}_{\bfi\sigma} \hc^{\phantom{\dagger}}_{\bfj\sigma} + 
       \hc^{\dagger}_{\bfj\sigma} \hc^{\phantom{\dagger}}_{\bfi\sigma} \right)
       + U\sum_\bfi \hn_{\bfi\uparrow}\hn_{\bfi\downarrow},
   \label{eq:Hamiltonian}
\end{align}
where $\hc^\dag_{\bfi\sigma}$ ($\hc_{\bfi\sigma}$) are the spin-$\sigma$ ($\sigma=\upa$, $\dna$) electron creation 
(annihilation) operator at site $\bfi$. $U > 0$ is the on-site Coulomb repulsion. $t$ is the 
hopping integral between two near-neighbor sites $\bfi$ and $\bfj$. Throughout this work, we
use $t$ as the unit of energy and set $t=1$. Due to the local nature of 
the Hubbard interaction, we choose to work with the spin-decomposition HS transformation\cite{Hirsch:1983}
in the calculations.

Fig.~\ref{fig:lattices} depicts the geometries of several $L_1 \times L_2$ lattices considered in 
our calculations. 
Here $L_1$ and $L_2$ are the linear dimension along the Bravais lattice vector $\bfa_1$ and $\bfa_2$ respectively. 
Both bipartite and geometrically frustrated lattices are considered, and periodic boundary conditions 
are assumed in all simulations.

% ------------------- figure ----------------------
\begin{figure}[b]
\includegraphics[scale=0.44]{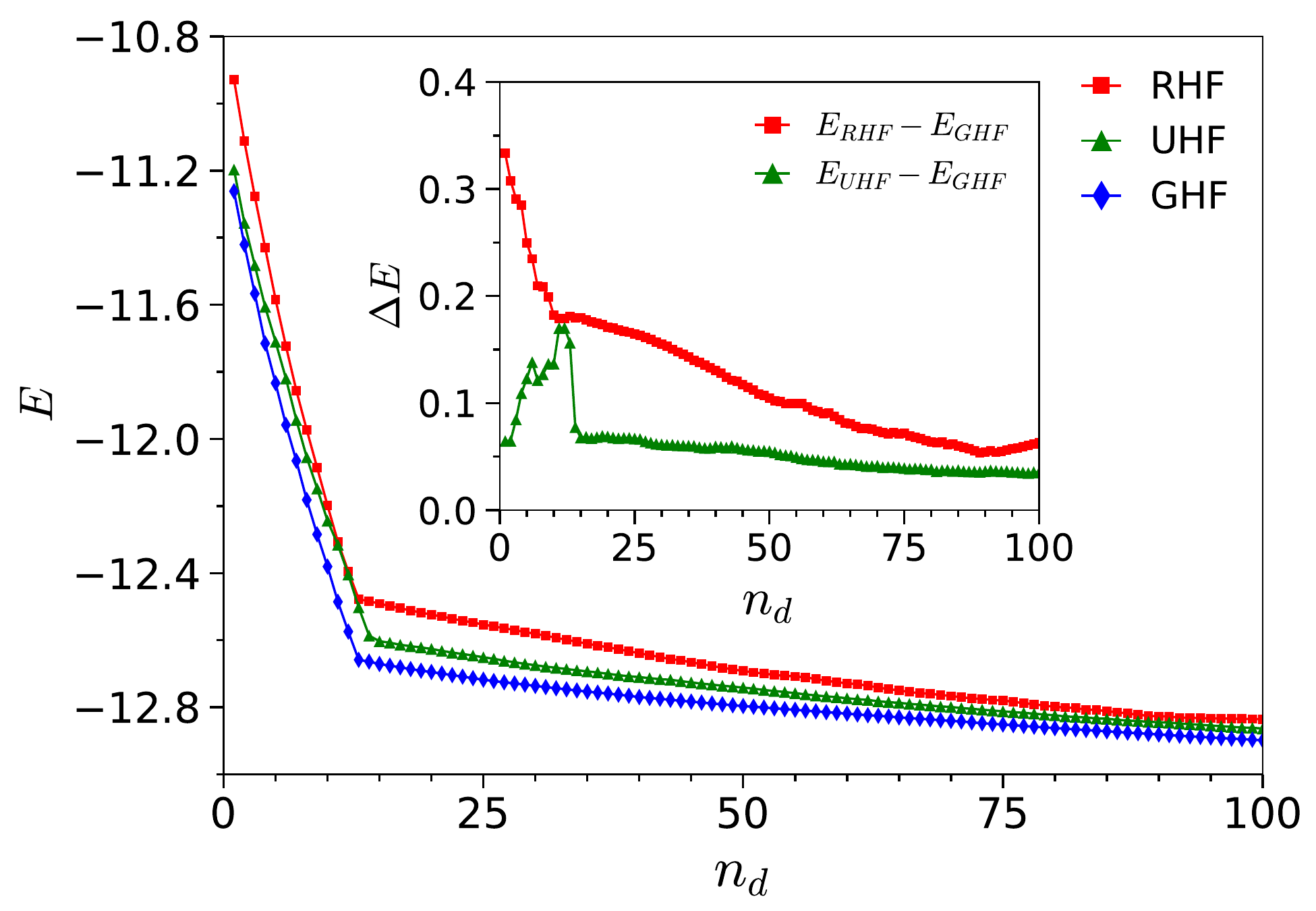}
\caption{Variational energy of RHF, UHF and GHF states versus the number versus the number of 
determinants $n_d$ for the half-filled $4\times 3$ triangular lattice at $U=4$. The calculations
are carried out using the FED theory.
The inset shows the energy of RHF and UHF states at a given number of determinants $n_d$ measured from 
the that of the corresponding GHF state. Clearly, $\Delta E > 0$ indicates that the GHF state 
has a lower variational energy.
}
\label{fig:GHF.vs.RHFUHF}
\end{figure}

\subsection{Results}

We begin by making two remarks. Firstly, we note that RHF or UHF orbitals can be considered as
special cases of the GHF wave function with vanishing off-diagonal spin components 
($\ca{}{\upa}\de{}{\dna}$ or $\ca{}{\dna}\de{}{\upa}$). To demonstrate the advantage of using the
GHF state at the variational level,\cite{Hammes-Schiffer:1993}
Fig.~\ref{fig:GHF.vs.RHFUHF} gives a simple comparison for the half-filled Hubbard triangular 
lattice at $U=4$. The figure shows that the GHF wave function clearly has the lowest variational
energy regardless of $n_d$.

\begin{figure}[b]
\includegraphics[scale=0.42]{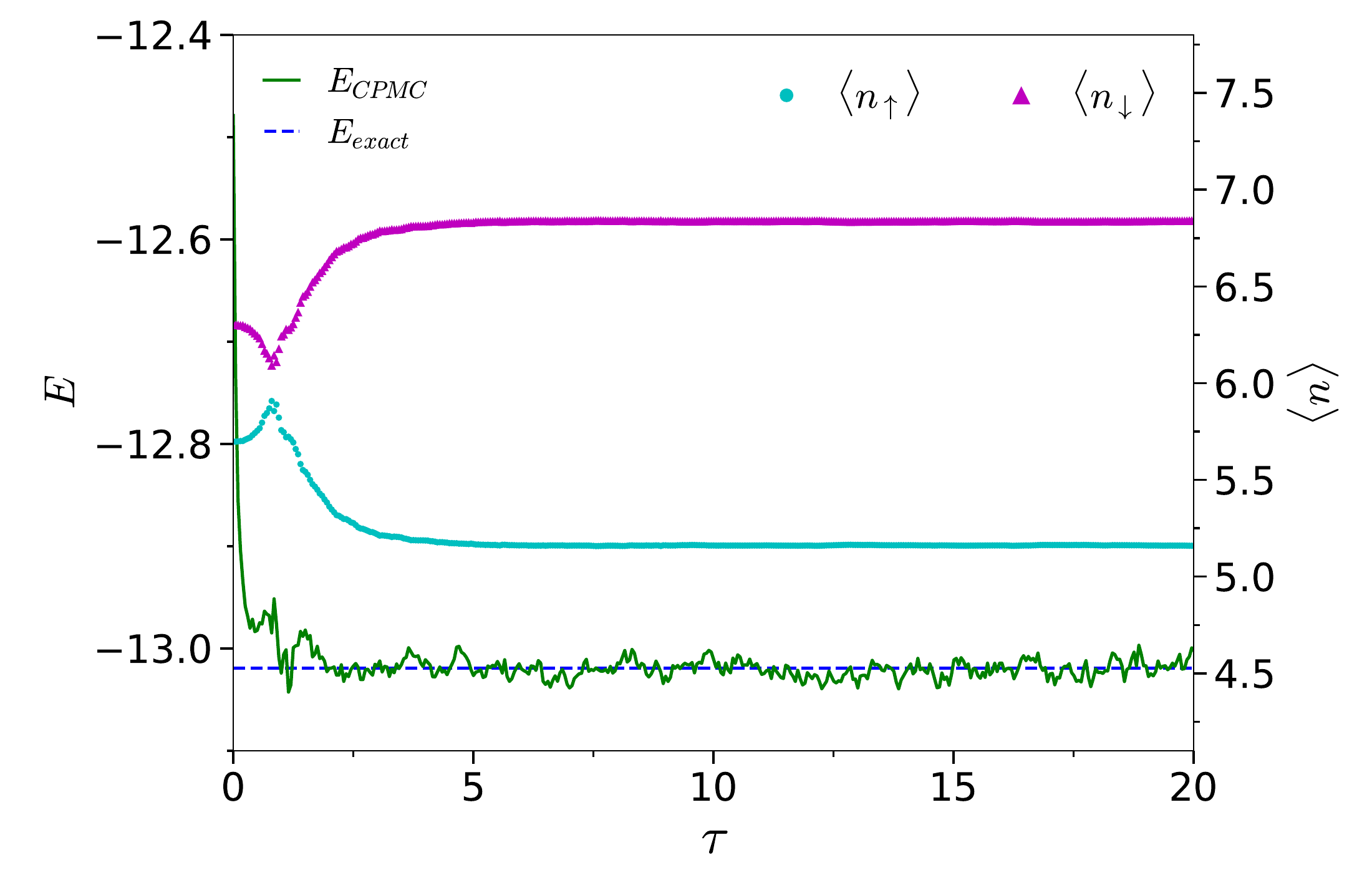}
\caption{Local energy and particle number as a function of projection time $\tau$.
Our system is a half-filled $4\times 3$ triangular lattice with $U=4$. The Trotter step
is $\Delta\tau=0.05$.
The CPMC energy is represented by the (green) solid line. The reference exact energy is
the (blue) dashed line. Density of the spin up and down electron is denoted by (cyan)
circle and (magenta) triangle respectively. In this simulation, we use a GHF orbital
to initialize the walkers. 
The trial state is a 10-determinant ResHF-optimized GHF wave function.
Note that, for each spin component, the particle number clearly deviates from the 
expected value $\ob{n}=6$.
}
\label{fig:GHFwalker}
\end{figure}

Secondly, while the total number of electrons in a GHF wave function is conserved, the particle number 
for each spin component $\ob{\hn_\upa}$ and $\ob{\hn_\dna}$ does not because spin-flip terms are now
explicitly included in the GHF orbitals. Therefore, simulations using GHF-type walkers will typically 
break the symmetry between $\ob{\hn_\upa}$ and $\ob{\hn_\dna}$. This is demonstrated by Fig.~\ref{fig:GHFwalker} 
which depicts $\ob{\hn_\upa}$ and $\ob{\hn_\dna}$ as well as the CPMC energy as a function of
imaginary projection time $\tau$ for the half-filled $4\times 3$ Hubbard triangular lattice. In this 
example, a 10-determinant GHF trial wave function is adopted. Walkers are initialized by a single-determinant GHF 
wave funcion. It can be clearly seen that although the energy converges reasonably well to the exact energy 
(apart from the Trotter error), the density does not. Therefore throughout this work, we will initialize 
random walkers using free-electron wave functions\footnote{In the case of open-shell fillings, we have implemented 
a small twist boundary condition $(\sim {\cal O}(10^{-3}))$ to break the degeneracy.} in which 
$\ob{\hn_\upa}$ and $\ob{\hn_\dna}$ are fixed at the desired value.

\begin{figure}[t]
\includegraphics[scale=0.45]{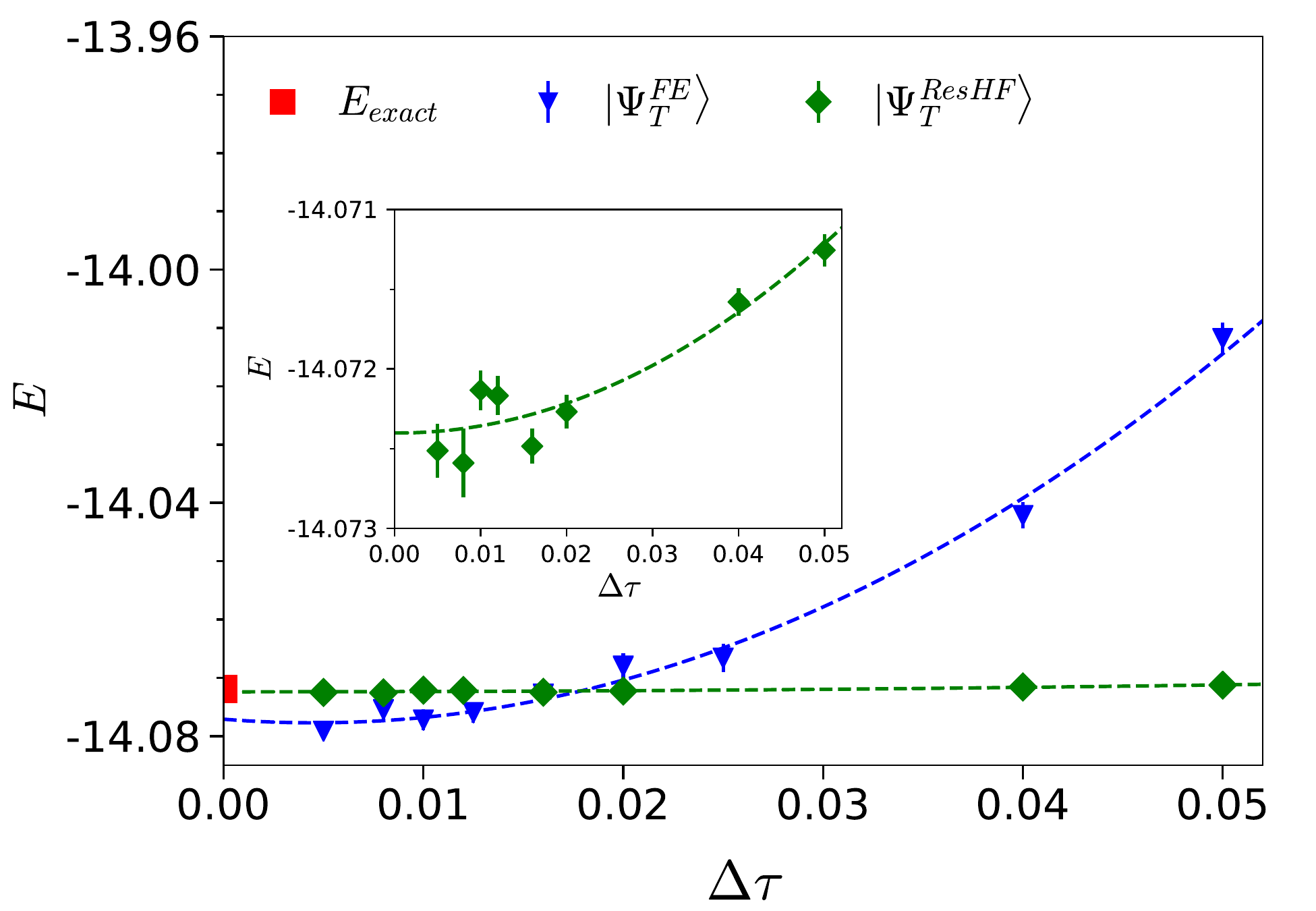}
\caption{Illustration of Trotter error correction. Simulations are carried out on the 
$2\times 2$ kagome lattice doped with 4 holes at $U=8$. Results obtained using
free-electron (FE) and 200-determinant ResHF-optimized GHF trial wave functions are
denoted by (blue) triangle and (green) diamond respectively. The (red) square marks the
exact energy. The inset shows detailed time-step dependence of the
QMC ground state energy for the ResHF-optimized GHF state.
}
\label{fig:Trotter}
\end{figure}

In Fig.~\ref{fig:Trotter}, we choose to demonstrate typical Trotter approximation behaviors in our calculations.
The sample system is a 4-hole doped $2\times 2$ kagome lattice at $U=8$. Since this is a closed-shell 
filling, we compare the $\dtau$-dependence of a free-electron (FE) trial wave function to that of a 200-determinant 
GHF state optimized using the ResHF theory (dubbed ResHF-optimized).
The multi-determinant trial wave function clearly has a weaker time-step dependence. Similar comparisons 
are also observed in other simulations. The relatively weaker $\dtau$-dependence seems to be typical in calculations
using multi-determinant trial states.\cite{Chang:2016} In this example, the extrapolated energy is lower than the 
exact data for the FE trial wave function. In other words
the mixed estimator $E=\ob{\Psi_T|\hH|\Phi}/\ob{\Psi_T|\Phi}$ used to compute the energy is not always variational.
Unlike the real-space fixed-node method,\cite{Foulkes:2001} random walkers in CPMC are represented by 
over-complete non-orthogonal Slater determinants. As a result, walkers removed by the constraint are 
not necessarily orthogonal to the remaining walkers, breaking the equivalence between the variational 
and mixed estimators. For detailed discussions and proposals of constructing variational energy estimators, 
we refer the readers to Ref.~\onlinecite{Carlson:1999}. In the present study we will not address the issue.

% ------------------- figure ----------------------

% ------------------- figure ----------------------

\begin{figure}[t]
\includegraphics[scale=0.43]{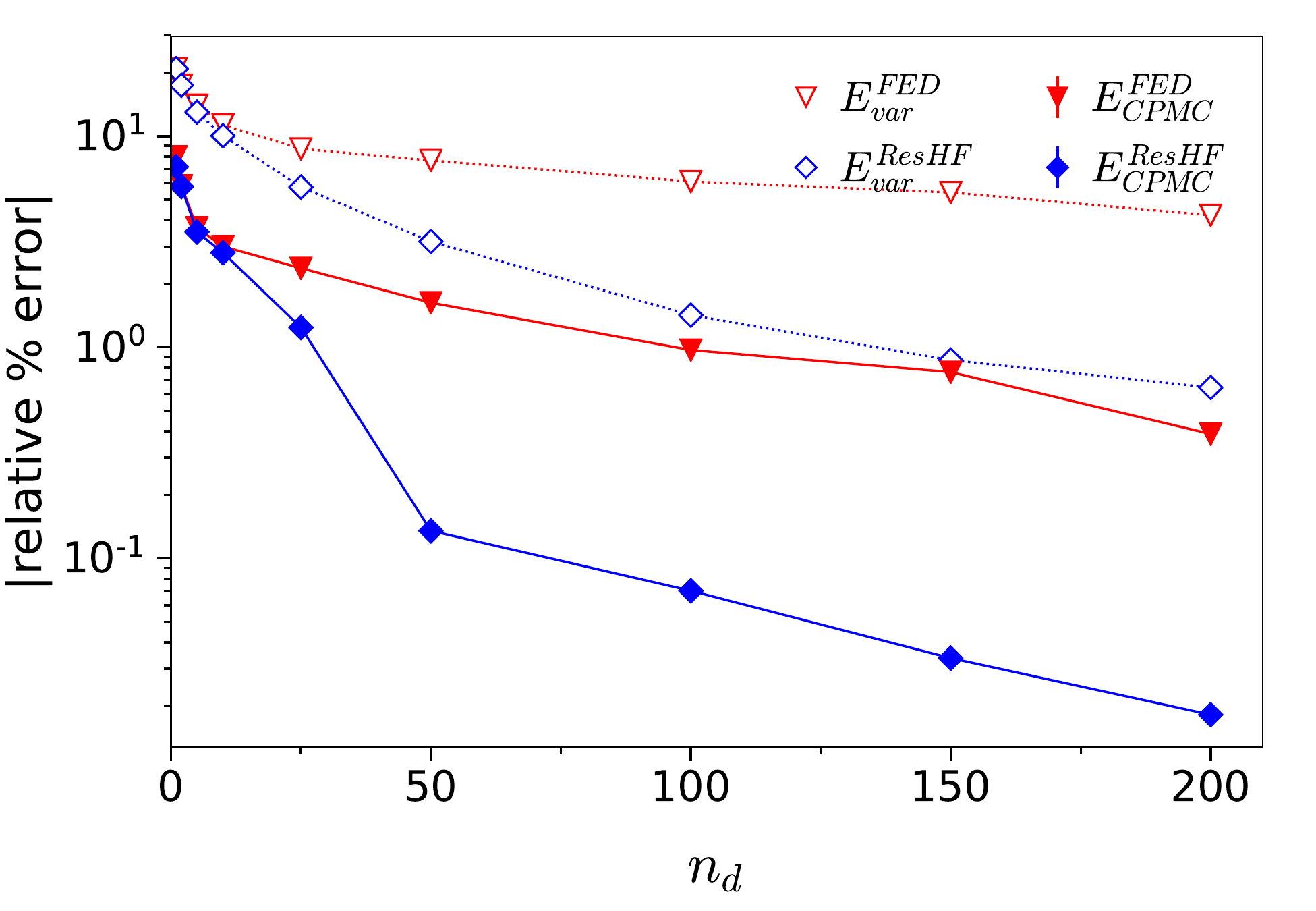}
\caption{
Accuracy comparison for the variational and CPMC results obtained using FED and ResHF-optimized
GHF wave functions. The system under consideration is the half-filled $2\times 2$ kagome lattice at $U=8$. 
}
\label{fig:FED.vs.ResHF}
\end{figure}

As mentioned in Section \ref{sec:Method}, we use a two-step approach to generate GHF trial wave functions. 
For a given $n_d$, a FED GHF wave function is constructed. The ResHF theory is applied subsequently to 
optimize all determinants simultaneously. In principle, both FED and ResHF-optimized GHF states can serve
the role of trial state. To examine their performance, we first compare the deviation of their 
variational energy from the exact ground state energy in Fig.~\ref{fig:FED.vs.ResHF}. The system under 
consideration is the half-filled $2\times 2$ kagome lattice at $U=8$.
Indeed, by adding more configurations to the multi-determinant expansion, the relative error in variational 
energy can be reduced systematically for both type of wave functions. In this particular example, however, 
one is able to gain a substantial improvement over the FED state by applying the ResHF theory. For example, at 
$n_d=200$, the deviation (defined as $|E-E_{ex}|/|E_{ex}|\times 100\%$, where $E_{ex}$ is the exact energy)
of the ResHF variational energy is $\sim 0.64\%$ while that of the FED variational energy is about $4.22\%$.
Similar comparisons are also observed in other lattice geometries. 

This comparison between FED and ResHF-optimized wave functions carries over to CPMC calculations. 
Also in Fig.~\ref{fig:FED.vs.ResHF}, results of a series of CPMC simulations using FED and ResHF-optimized 
GHF trial wave functions are plotted for $n_d$ up to 200 determinants. Both calculation results are 
approaching the exact energy monotonically as more determinants are included and optimized. At $n_d=200$, 
the FED trial wave function gives $E_{CPMC} = -5.60252(28)$, corresponding to a relative error $\sim 0.388\%$.
Using the ResHF-optimized GHF state as the trial wave function, the CPMC ground state energy is 
$E=-5.62334(11)$, which is $0.018\%$ (in relative error) away from the exact energy. 

We would like to mention that in our benchmark calculations (see Table \ref{tbl:benchmark}), there are 
several closed-shell hole-doped systems under periodic boundary conditions where both FED and ResHF-optimized 
states give almost comparable results. Furthermore, we found in our benchmark data that there are situations
where the CPMC energy does not improve systematically. It turns out that these exceptions are caused by poor 
FED determinants. Therefore, as in any numerical mean-field calculations, it is necessary to use different 
initial configurations and make sure the FED result converge properly. When care is taken, the combined FED
and ResHF theory can produce high quality trial states for Monte Carlo simulations.

% ------------------- figure ----------------------

We list detailed benchmark calculations results in Table \ref{tbl:benchmark}.
In each test case, we have used both FED and ResHF-optimized trial wave functions consisting of as many as 
200 determinants. In the table $n_d$ represents the number of ResHF-optimized determinants that gives the 
best agreement with the exact diagonalization result. Energies reported in the table have been extrapolated 
to the $\dtau\rightarrow 0$ limit following the example of Fig.~\ref{fig:Trotter}.

The table shows the advantage of optimizing FED determinants for both mean-field and many-body calculations.
Although in a few instances the FED and ResHF-optimized states appear to have the same variational energy,
the latter still leads to a better agreement in Monte Carlo calculations.
Generally speaking, at weak couplings $U < W$, where $W$ is the bandwidth of the tight-binding Hamiltonian, 
the data suggest that a handful number of determinants on the order $n_d\sim {\cal O}(10)$ is sufficient to 
reduce the constrained-path systematic error in the calculation results regardless of the geometries. 
On half-filled bipartite lattices, in particular, a single-determinant GHF trial state is able to generate 
essentially exact results for the square and honeycomb lattices at $U=4$, a situation that is consistent with 
previous reports.\cite{Qin:2016} 
At $U=8$ ($U\sim W$) the number of determinants that gives the best agreement goes up to the order 
${\cal O}(10^2)$. This behavior is to be expected since, at the ansatz wave function level, strong 
correlations (in the sense that $U \gtrsim W$) typically require multi-determinantal wave functions 
to capture correlation effects.\cite{Clark:2011,Morales:2012} For extremely correlated systems where $U > W$, 
it is very likely that more than ${\cal O}(10^2)$ are required for the result to converge. A possible 
alternative option would be using the Gutzwiller projector, an approach shown to be quite effective
at large couplings.\cite{Chang:2016} However, we will leave the construction of trial states by 
Gutzwiller-projecting GHF determinants for future investigations.

Finally, it is worth mentioning that on the slightly doped square lattice, competing phases such as incommensurate
magnetic ordering, charge inhomogeneity, and possibly $d$-wave superconducting state may exist and compete 
with each other. Without restoring the symmetry of the exact ground state in the trial wave functions, the 
Monte Carlo results appear to have relatively larger deviations, as can be observed in Table \ref{tbl:benchmark}.
Apart from this challenging scenario, the benchmark data generally suggest that the combined FED and ResHF 
theories is able to generate high quality {\it symmetry-broken} trial wave functions that can much reduce 
the systematic errors caused by the constrained-path approximation.

\section{Summary}

In this work, we have studied the performance of multi-determinant generalized Hartree-Fock trial wave
function in CPMC calculations. By comparing the results with exact eigenenergies available on small
clusters, it is shown that the optimized multi-determinant GHF state is able to produce satisfactory
benchmark results {\it without} explicitly restoring the symmetries of the Hamiltonian at a manageable
computational cost. The quality of the multi-determinant GHF state can be systematically improved by
including more determinants in the FED and ResHF calculations. The combined (symmetry-broken) FED-ResHF 
GHF approach thus promises a flexible and straightforward technique for accurate Monte Carlo calculations.

\section{Acknowledgement}
We would like to thank Gustavo Scuseria and Carlos Jim\'enez-Hoyos for important discussions. We also
thank Carlos Jim\'enez-Hoyos for providing us the access of the PHF program.
C.-C. C would like to thank Yao Wang (Stanford University) for providing exact diagonalization 
data for the $4\times 4$ triangular lattice.
This work was performed under the auspices of the U.S. Department of Energy by Lawrence 
Livermore National Laboratory under Contract DE-AC52-07NA27344, 15-ERD-013.

\begin{table*}
\begin{ruledtabular}
\begin{tabular}{@{}cccrrrrrrrrrr@{}}
          & $\conf$ & $U$ & $n_d$ & $E_{ex}$   & $\EvFED$   & $\dvFED$  & $\EvRes$   & $\dvRes$  & $\EcFED$       & $\dqFED$ &  $\EcRes$        & $\dqRes$   \\
  \hline                                                                                                                                                    
  $\sysa$ & $(5,5)$ &  4  &   1   & -19.58094  &  -17.75000 &   9.35   &  -17.75000 &   9.35   &  -19.58085(62) &  0.0004  &   -19.58085(62)  &  0.0004    \\
          &         &  8  &   25  & -17.51037  &  -15.65947 &  10.57   &  -16.10752 &   8.01   &  -17.50538(41) &  0.0285  &   -17.51059(51)  &  0.0012    \\
          & $(6,6)$ &  4  &   10  & -17.72958  &  -16.63047 &   6.19   &  -16.91423 &   4.59   &  -17.74224(70) &  0.0714  &   -17.72972(18)  &  0.0007    \\
          &         &  8  &  150  & -14.92531  &  -13.82672 &   7.36   &  -14.20179 &   4.84   &  -14.74923(60) &  1.1797  &   -14.86261(53)  &  0.4200    \\
          & $(7,7)$ &  4  &   2   & -15.74459  &  -14.28826 &   9.24   &  -14.34519 &   8.88   &  -15.74299(66) &  0.0101  &   -15.74432(62)  &  0.0017    \\
          &         &  8  &  200  & -11.86884  &  -11.43362 &   3.66   &  -11.45992 &   3.44   &  -11.83087(27) &  0.3199  &   -11.84089(17)  &  0.2354    \\
          & $(8,8)$ &  4  &   1   & -13.62185  &  -12.56655 &   7.74   &  -12.56655 &   7.74   &  -13.62135(97) &  0.0036  &   -13.62135(97)  &  0.0036    \\
          &         &  8  &   25  &  -8.46888  &   -8.08078 &   4.58   &   -8.17569 &   3.46   &   -8.08492(33) &  4.5337  &    -8.46869(21)  &  0.0022    \\
  \hline                                                                                                                        
  $\sysb$ & $(4,4)$ &  4  &   25  & -11.68142  &  -11.44979 &   1.98   &  -11.55725 &   1.06   &  -11.66498(22) &  0.1407  &   -11.68099(21)  &  0.0036    \\
          &         &  8  &   200 & -10.39861  &  -10.21772 &   1.73   &  -10.35093 &   0.45   &  -10.36095(30) &  0.3621  &   -10.39890(23)  &  0.0027    \\
          & $(5,5)$ &  4  &   2   & -11.14382  &   -9.80815 &  11.98   &   -9.84599 &  11.64   &  -11.13732(57) &  0.0583  &   -11.14386(51)  &  0.0003    \\
          &         &  8  &   50  &  -8.65004  &   -7.59496 &  12.19   &   -8.24504 &   4.68   &   -8.12760(77) &  6.0397  &    -8.64680(36)  &  0.0374    \\
          & $(6,6)$ &  4  &   1   &  -9.59093  &   -7.97626 &  16.83   &   -7.97626 &  16.83   &   -9.59011(51) &  0.0085  &    -9.59011(51)  &  0.0085    \\
          &         &  8  &   10  &  -5.36058  &   -4.87161 &   9.12   &   -4.92776 &   8.07   &   -5.36036(24) &  0.0041  &    -5.36043(27)  &  0.0027    \\
  \hline                                                                                                                        
  $\sysc$ & $(4,4)$ &  4  &   25  & -17.69611  &  -17.33308 &   2.05   &  -17.49831 &   1.11   &  -17.69799(5)) &  0.0104  &   -17.69786(6)   &  0.0098    \\
          &         &  8  &  200  & -15.74457  &  -15.54768 &   1.25   &  -15.72645 &   0.11   &  -15.72682(29) &  0.1127  &   -15.74427(10)  &  0.0019    \\
          & $(5,5)$ &  4  &  2    & -16.70493  &  -15.42878 &   7.63   &  -15.44900 &   7.51   &  -16.70078(69) &  0.0248  &   -16.70402(45)  &  0.0054    \\
          &         &  8  &  150  & -13.57065  &  -13.20282 &   2.71   &  -13.45542 &   0.84   &  -13.56406(36) &  0.0485  &   -13.56971(26)  &  0.0069    \\
          & $(6,6)$ &  4  &  5    & -13.01919  &  -11.69009 &  10.20   &  -11.74491 &   9.78   &  -13.00762(51) &  0.0888  &   -13.01510(35)  &  0.0314    \\
          &         &  8  &  200  &  -6.99127  &   -6.36472 &   8.96   &   -6.84717 &   2.06   &   -6.60580(66) &  5.5135  &    -6.97816(11)  &  0.1875    \\
  \hline                                                                                                                        
  $\sysd$ & $(7,7)$ &  4  &  10   &  -25.6558  &  -23.75000 &   7.42   &  -23.75000 &   7.42   &   -25.4457(13) &  0.8189  &    -25.6598(15)  &  0.0155    \\ 
          &         &  8  &  50   &  -18.9764  &  -16.64556 &  12.28   &  -16.69779 &  12.00   &   -18.9681(19) &  0.0437  &    -18.9798(19)  &  0.0179    \\
  \hline                                                                                                                        
  $\syse$ & $(4,4)$ &  4  &  2    & -16.05774  &  -14.66666 &   8.66   &  -14.66666 &   8.66   &  -16.06150(29) &  0.0234  &   -16.05477(80)  &  0.0184    \\
          &         &  8  &  25   & -14.07191  &  -13.13592 &   6.65   &  -13.34475 &   5.16   &  -14.07003(24) &  0.0133  &   -14.07219(29)  &  0.0019    \\
          & $(5,5)$ &  4  &  5    & -13.98637  &  -12.99466 &   7.09   &  -13.15281 &   5.95   &  -13.98346(69) &  0.0208  &   -13.98673(76)  &  0.0025    \\
          &         &  8  &  150  & -11.00945  &  -10.13409 &   7.95   &  -10.81013 &   1.81   &  -10.62814(28) &  3.4634  &   -11.01117(32)  &  0.0156    \\
          & $(6,6)$ &  4  &  5    & -10.57279  &   -9.44590 &  10.65   &   -9.60950 &   9.11   &  -10.57687(27) &  0.0385  &   -10.57193(64)  &  0.0081    \\
          &         &  8  &  200  &  -5.62437  &   -5.38700 &   4.22   &   -5.58810 &   0.64   &   -5.60252(28) &  0.3884  &    -5.62334(11)  &  0.0183    \\
 %  \botrule
\end{tabular}
\end{ruledtabular}
\caption{Comparison of ground state energy of the 2D Hubbard model. In the first column, `sqr', `hc', `tri', and `kag' 
are short names of square, honeycomb, triangular and kagome lattices. $\conf$ denotes the electronic configuration,
$n_d$ is the number of GHF determinants in the wave function expansion. $E_{ex}$ lists the exact diagonalization data. 
The column $\EvFED$ and $\EvRes$ lists variational energy of the FED and ResHF-optimized GHF wave functions respectively. 
$\EcFED$ and $\EcRes$ are the CPMC energy obtained using corresponding GHF trial wave functions. The symbol $\Delta$
shows percentage relative error defined as $\Delta=|E-E_{ex}|/|E_{ex}|\times 100\%$, where $E$ is the variational or CPMC 
energy.
The exact energy of the $4\times 4$ square and triangular lattices is obtained from
Ref.~\onlinecite{Dagotto:1992} and \onlinecite{Wang:2017} respectively. The rest of the exact data are obtained 
from using the ALPS library.\cite{Bauer:2011}
}
\label{tbl:benchmark}
\end{table*}

\bibliography{reference}

\end{document}